\documentclass[conference,compsoc]{IEEEtran}
\newcommand{\xiao}[1]{\textcolor{black}{{#1}}}

\usepackage{listings}

\usepackage{xcolor}
\usepackage[most]{tcolorbox}

\definecolor{takeawaybg}{RGB}{245,245,245}
\definecolor{takeawayline}{RGB}{90,90,90}

\newtcolorbox{takeawaybox}{
  enhanced,
  width=\linewidth,
  colback=takeawaybg,
  colframe=takeawaybg,
  boxrule=0pt,
  borderline west={1.5pt}{0pt}{takeawayline},
  left=10pt,
  right=10pt,
  top=7pt,
  bottom=7pt,
  before skip=6pt,
  after skip=6pt,
  sharp corners,
  breakable
}
\newcommand{\TakeawayBox}[1]{%
  \begin{takeawaybox}
    #1
  \end{takeawaybox}%
}

\newtcolorbox{promptbox}[2][]{
  enhanced,
  breakable,
  colback=white,
  colframe=black!70,
  colbacktitle=black!70,
  coltitle=white,
  title={#2},
  fonttitle=\bfseries,
  left=6pt, right=6pt, top=6pt, bottom=6pt,
  boxrule=0.6pt,
  arc=2mm,
  #1
}

\ifCLASSOPTIONcompsoc
  \usepackage[nocompress]{cite}
\else
  \usepackage{cite}
\fi
\ifCLASSINFOpdf
  \usepackage{graphicx}
\else
\fi
\usepackage{url}

\usepackage[table]{xcolor}
\usepackage{booktabs}
\usepackage{hyperref}
\usepackage{amssymb}
\usepackage{amsmath}
\usepackage{enumitem}
\usepackage{multirow}
\usepackage{tikz,xcolor}
\definecolor{LightBlue}{HTML}{F0F8FF}
\definecolor{SectionGray}{HTML}{ECEFF1}

\begin{document}
%
\title{AudioTrace: File-Level Membership Detection for Audio Large Language Models}

\author{
\IEEEauthorblockN{
Guangzhi Sun\textsuperscript{1,\dag},
Xiao Zhan\textsuperscript{1,\dag},
Shuai Wang\textsuperscript{2,4},
Jose Such\textsuperscript{3} and
Phil Woodland\textsuperscript{1}
}
\IEEEauthorblockA{
\textsuperscript{1}University of Cambridge, Cambridge, United Kingdom\\
\textsuperscript{2}School of Intelligence Science and Technology, Nanjing University, Suzhou, P.R. China\\
\textsuperscript{3}INGENIO (CSIC-Universitat Politècnica de València), Valencia, Spain\\
\textsuperscript{4}State Key Lab. of Novel Software Technology, Nanjing University, Nanjing, P.R. China\\
\textsuperscript{\dag}These authors contributed equally.\\
Emails: \{gs534, xz603, pw117\}@cam.ac.uk, shuaiwang@nju.edu.cn, jose.such@csic.es
}
}

\maketitle

\begin{abstract}
Significant privacy and copyright concerns are raised by the use of publicly available audio recordings to train audio large language models (LLMs), yet effective auditing methods remain limited. Existing membership inference attacks typically assume knowledge of complete training examples, including their associated textual supervision. In practice, however, audio recordings may be paired with privately generated transcripts, captions, or question-answer pairs that are unavailable to owners of the recordings. This motivates audio-file-level membership inference, which determines whether a recording was used for training without assuming knowledge of its training tasks and annotations. This paper provides the first systematic investigation of this challenge and introduces AudioTraceBench, a controlled benchmark comprising 1,285 audio recordings with explicit fine-tuning membership labels. Evaluations across three audio LLMs show that existing text- and video-based membership inference methods adapted to audio perform close to random guessing.

To address this challenge, AudioTrace is proposed, where audio-conditioned proxy questions are generated using the target model, and membership evidence is extracted from its responses and aggregated at the recording level. Consistent improvements over existing baselines are observed, with AUROC reaching 0.60 and increasing further under increased fine-tuning exposure or adaptation capacity. Further experiments were conducted to examine the factors influencing membership signals for audio-file-level detection. These findings establish audio-file-level membership inference as a distinct and practically important auditing challenge, with AudioTrace providing an effective initial approach to this challenge.

\end{abstract}


%
\IEEEpeerreviewmaketitle

\section{Introduction}

Audio large language models (Audio LLMs) are increasingly capable of understanding and reasoning over audio inputs, enabling applications such as speech recognition, audio captioning, and audio question answering~\cite{salmonn,qwen2audio,salmonn2,stepaudio2,qwen25omni
}. However, training Audio LLMs on large-scale audio collections raises significant privacy and copyright concerns. Audio recordings may contain identifiable voices, private conversations, and copyrighted performances, making their inclusion in training datasets potentially sensitive. This raises a fundamental question for individuals and content owners: \textbf{Can they determine whether their audio recordings were used to train an audio LLM?}


Membership inference attacks (MIAs) offer a potential approach to this auditing problem by determining whether a given sample was used to train a target model~\cite{shokri2017membership}. 
In text-based LLMs, membership inference is operated on individual text sequences using model confidence signals~\cite{shi2024detecting,zhang2025min}, 
Recent studies have extended MIAs to multimodal models.
In particular, video-file-level methods exploit temporal dependencies between frames to infer whether a video was included in training~\cite{li2025vid,videomia}. However, directly transferring these approaches to Audio LLMs remains challenging. Unlike videos, which can be represented as sequences of visually interpretable frames, audio recordings contain continuous, often overlapping acoustic events, making temporal structure alone an uncertain indicator of membership. More importantly, Audio LLMs are often fine-tuned using publicly available recordings, such as podcasts and YouTube videos, paired with task-specific annotations created by model developers, which are typically not accessible to the owner of the audio file. This makes determining whether a particular audio file \xiao{(file-level)} was used for training substantially more challenging than inferring membership of a known audio-question-answer pair \xiao{(sample-level)}. 
To address this challenge, we present, to the best of our knowledge, the first systematic study of \textbf{audio-file-level membership inference against Audio LLMs}, aiming to determine whether a recording was used for fine-tuning \textbf{without knowledge of its associated training queries and annotations}.

To assess the feasibility of audio-file-level membership inference, existing MIA methods developed for text and video models are first adapted and evaluated in this setting. To enable a controlled evaluation, AudioTraceBench, a benchmark of 1,285 speech-centric audio recordings collected from YouTube videos published after January 1, 2026, is constructed. News and conversational content containing independently timestamped events is prioritised to reduce the risk of overlap with pretraining data. Three Audio LLMs, including Qwen2.5-Omni, SALMONN-2, and Step-Audio-2, were fine-tuned from pre-2026 checkpoints \xiao{(i.e., checkpoints released before January 1, 2026)} on a controlled subset of the benchmark to establish ground-truth membership labels. 
The results show that existing text- and video-based MIA methods, when adapted to audio-file-level auditing, perform close to random guessing, highlighting the difficulty of audio-file-level membership inference and the need for a dedicated approach.

Motivated by these findings, AudioTrace is proposed as a proxy-question-based method for audio-file-level membership inference. Since membership evidence can depend strongly on the probing instructions~\xiao{\cite{wen2024membership}}, a key challenge is identifying informative questions without knowing the original fine-tuning details. AudioTrace addresses this challenge by leveraging the target model's learned audio-instruction associations to generate supervision-aligned proxy questions. AudioTrace first prompts the target model to generate questions about a candidate recording, then queries the same model to produce corresponding answers. Membership evidence is extracted from the generated answers using token-level statistics, and the strongest question-level signals are aggregated through top-$K$ averaging to obtain a recording-level membership score.  
\xiao{Experiments across three Audio LLMs demonstrate consistent improvements over adapted baselines.} Furthermore, comparisons with known training question-answer pairs reveal a substantial performance gap, highlighting the difficulty of audio-file-level membership inference under unknown training annotations.


\noindent \textbf{Contributions.}

\begin{itemize}[leftmargin=8pt]
    \item We formulate and investigate \textbf{audio-file-level membership inference} against Audio LLMs, where the adversary seeks to determine whether a recording was used during fine-tuning without knowing its associated training queries and annotations. The \textbf{substantial difficulty} of this setting is demonstrated by comparing with membership inference using known training question-answer pairs.

    \vspace{0.1cm}
    \item We introduce \textbf{AudioTraceBench}, a benchmark of 1,285 recordings from post-2026 YouTube videos. Using three pre-2026 audio LLM checkpoints, controlled fine-tuning membership labels were established, validated against blind baselines, and adapted to representative text- and video-based MIAs for audio-file-level evaluation.

    \vspace{0.1cm}
    \item 
    We propose \textbf{AudioTrace}, a proxy-question-based method for membership inference without training annotations. Experiments show consistent improvements over baselines and reveal key factors influencing membership signals.
\end{itemize}

\section{Related Work}

\subsection{Membership Inference for LLMs}

Membership inference attacks (MIAs) aim to determine whether a given sample was included in a model's training set. Early studies introduced shadow-model attacks, loss-based tests, and likelihood-ratio methods for supervised models~\cite{shokri2017membership,yeom2018privacy,carlini2022membership}. These approaches were subsequently extended to language models, where membership is commonly inferred from token-level losses, probabilities, and other output-distribution statistics~\cite{song2019auditing,mireshghallah2022quantifying}. Recent methods include Min-K\% Prob and Min-K\%++, perturbation-based calibration, and contextual or relative-likelihood approaches~\cite{shi2024detecting,zhang2025min,galli2024noisy,wang2025recall}.

Membership inference has also been extended to multimodal models~\cite{hu2022m,li2024membership,li2025vid,dong2026membership}. Existing approaches exploit output-reference similarity, token-level confidence, and entropy-based statistics to infer membership of multimodal inputs~\cite{hu2022m,li2024membership}. Vid-SME further investigates video membership inference by measuring changes in model confidence under temporal frame transformations~\cite{li2025vid}. For audio models, prior work has investigated speaker-level membership, ASR auditing, and the memorization of acoustic-text associations~\cite{chen2023slmia,teixeira2024improving,wang2024unintended,jagielski2024noise}. Recent studies on Audio LLMs examine membership signals using audio-text alignment and prediction statistics such as perplexity, Min-K\%, and entropy~\cite{cheng2025gibberish,dong2026membership}. Other work investigates black-box auditing of generative music models and dataset-level inference from collections of audio recordings~\cite{liu2026auditing,proboszcz2025membership}.
\\
\noindent \textbf{Research Gap.} Existing studies have investigated membership inference at the level of training examples, speakers, or entire datasets, but have not systematically addressed audio-file-level membership inference without knowledge of the original training queries and annotations. \xiao{To address this gap, we introduce AudioTraceBench, a benchmark for controlled evaluation, and AudioTrace, a novel method for determining whether an audio recording was used for training using only the audio.} 

\subsection{Privacy and Copyright Auditing}
\label{sec:}

Beyond its role as an attack, membership inference provides a potential tool for privacy and copyright auditing. Prior work has demonstrated the memorization and exposure of sensitive training content, including personal information and acoustic-text associations~\cite{carlini2021extracting,wang2024unintended,dong2026membership}. Membership- and dataset-inference methods have also been explored for auditing copyrighted books, music, and proprietary datasets~\cite{meeus2024did,maini2024llm,proboszcz2025membership}. More broadly, data-use auditing must address uncertainty about model training and the reliability of the resulting evidence, motivating research on auditing with limited knowledge of the underlying learning task and on the reliability of membership inference as a privacy tool~\cite{huang2024general,wang2025membershiptools}. While dataset-level auditing provides evidence about the use of a collection, recording-level membership inference addresses the more specific question of whether an individual owner's audio was included in training. This distinction motivates auditing methods that can operate on individual recordings.

\section{Audio-file-level Membership Inference}

\subsection{Problem Formulation}
\label{sec:problem_formulation}

We study audio-file-level MIA under the supervised fine-tuning (SFT) paradigm, a common practice for adapting pretrained Audio LLMs to downstream audio-understanding tasks~\cite{salmonn,salmonn2,qwen2audio,videosalmonn,li2025vid}. This setting reflects practical scenarios in which model developers fine-tune publicly available Audio LLMs using audio paired with task-specific annotations, while allowing us to control the data used in fine-tuning for reliable evaluation explicitly. Formally, let $f_\theta$ denote an Audio LLM fine-tuned from a pretrained model $f_{\theta_0}$ on a dataset $\mathcal{D}_{\mathrm{train}} = \{(a_i, q_i, y_i)\}_{i=1}^{N}$, where $a_i$ is an audio recording, $q_i$ is a textual instruction, and $y_i$ is the corresponding response. The same recording may appear in multiple training examples with different instructions and responses, spanning tasks such as automatic speech recognition, audio captioning, and question answering.

Unlike conventional example-level MIAs, which infer whether a specific audio-instruction-response tuple was used for training, audio-file-level MIA aims to determine whether a given recording $a$ appeared in the fine-tuning dataset, regardless of its associated supervision. Formally, we define the membership indicator as
\begin{equation}
    m(a) = \mathbb{I}\left[
    \exists (q,y) \quad \text{such that} \quad (a,q,y)
    \in \mathcal{D}_{\mathrm{train}}
    \right].
\end{equation}
An attacker $\mathcal{A}$ is given a candidate recording $a$ and query access to the target fine-tuned model $f_\theta$, and seeks to predict $m(a)$. The attacker computes an audio-level membership score $s_{\mathcal{A}}(a;f_\theta)$ and makes a binary decision according to
\begin{equation}
    \hat{m}(a) =
    \mathbb{I}\left[
    s_{\mathcal{A}}(a;f_\theta) \geq \tau
    \right],
\end{equation}
where $\tau$ is a decision threshold. The objective is to distinguish member from non-member recordings using evidence obtained from the target model, without requiring knowledge of the specific instructions used during fine-tuning.

\subsection{Threat Model}
\label{sec:threat_model}

\noindent\textbf{Adversarial objective.}
We consider an external auditor, such as a content creator or recording owner, who seeks to determine whether a specific audio recording was used to fine-tune a target audio LLM. The adversary aims to infer the recording-level membership indicator $m(a)$, rather than the membership of a particular audio-instruction-response tuple. The attacker targets membership in supervised fine-tuning datasets used to adapt pretrained LLMs to downstream audio-understanding tasks.

\noindent\textbf{Adversarial knowledge.}
We assume the adversary possesses the candidate audio recording $a$ and knows the target model's audio-text interaction interface, such as associated Python packages to run inference. However, the adversary has no knowledge of the target's fine-tuning dataset, including the textual instructions and reference responses associated with $a$. In particular, the adversary cannot assume access to the original transcripts, captions, or question-answer pairs used during training. We also do not assume access to the pretrained checkpoint $f_{\theta_0}$ or other models sharing the target's training history.

\noindent\textbf{Adversarial capabilities.}
The adversary can repeatedly query the target model with a candidate recording and arbitrary textual instructions, and observe the generated responses. We distinguish two levels of model access:
\emph{(i) response-only access}, where the adversary observes only generated text; and
\emph{(ii) probability access}, where the adversary can additionally obtain token-level output probabilities or logits.
Response-only access supports behavioral attacks based on the model's predictions, while probability access enables likelihood- and entropy-based membership statistics.
AudioTrace operates under the latter setting, using the target model to generate proxy questions and score its responses without accessing model parameters or gradients.

\noindent\textbf{Evaluation scope.}
We evaluate this threat model under controlled supervised fine-tuning, where the inclusion of each candidate recording is explicitly determined by the experimenter. Membership labels and training annotations are used only to establish ground truth and evaluate attacks, and are not available to the adversary. We additionally consider access to the original training questions as a privileged comparison setting to quantify the difficulty introduced by unknown supervision.

\subsection{Direct Audio Membership Inference}
\label{sec:direct_audio_mia}

We first investigate whether existing membership inference methods developed for text and video can be adapted to audio-file-level auditing. We consider three families of approaches: (1) direct probability-based attacks, which infer membership from the model's output distributions; (2) Audio-SME, adapted from Vid-SME, which exploits the temporal ordering of audio segments using Sharma–Mittal entropy; and (3) an audio adaptation of VideoMIA, which leverages temporal reasoning over the chronological structure of audio recordings to infer membership. All methods operate directly on the candidate recording without requiring its original training instructions or reference responses.

\textbf{Direct probability-based attacks.}
Following the multimodal probability-extraction approach of Vid-SME~\cite{li2025vid}, we obtain the target model's output distributions at positions corresponding to the input audio representation. Specifically, given an audio recording $a$ which is to be converted to a sequence of vectors sent to the LLM backbone. Let $\mathcal{I}(a)$ denote the set of audio-associated positions in the model's input sequence. We extract
\begin{equation}
    p_t(v) = \operatorname{softmax}(z_t)_v,
    \quad t \in \mathcal{I}(a),
\end{equation}
where $z_t$ denotes the target model's output logits and $v$ indexes its output vocabulary. Let $T=|\mathcal{I}(a)|$ denote the number of selected positions.
Unlike text-based MIAs, these positions do not have corresponding ground-truth vocabulary labels. We therefore assign each position a pseudo-label $\tilde{y}_t = \arg\max_v p_t(v)$, enabling confidence-based adaptations of conventional token-level metrics.

We first consider perplexity (PPL) and Min-K\% probability~\cite{shi2024detecting}. PPL is computed as
\begin{equation}
    \operatorname{PPL}(a)
    = \exp\left(
        -\frac{1}{T}
        \sum_{t\in\mathcal{I}(a)}
        \log p_t(\tilde{y}_t)
    \right).
\end{equation}
For Min-K\%, we select the fraction $\rho$ of positions with the lowest pseudo-label log-probabilities and average their negative log-probabilities:
\begin{equation}
    S_{\mathrm{MinK},\rho}(a)
    = -\frac{1}{|\mathcal{T}_\rho|}
    \sum_{t\in\mathcal{T}_\rho}
    \log p_t(\tilde{y}_t),
\end{equation}
where $\mathcal{T}_\rho \subseteq \mathcal{I}(a)$ contains the selected positions.

We additionally evaluate entropy-based attacks, including Shannon entropy and R\'enyi entropy of different orders. For a distribution $p_t$, R\'enyi entropy is defined as
\begin{equation}
    H_\alpha(p_t)
    = \frac{1}{1-\alpha}
    \log \sum_v p_t(v)^\alpha,
\end{equation}
with Shannon entropy recovered as $\alpha \rightarrow 1$. We evaluate $\alpha\in\{0.5,1,2,\infty\}$ and aggregate the resulting position-wise entropy values using their mean or the Max-K\% statistic, which averages the largest fraction $\rho$ of entropy values. These methods yield recording-level membership scores directly from audio-associated output distributions.

\textbf{Audio-SME.}
We adapt Vid-SME~\cite{li2025vid} to audio recordings and refer to the resulting baseline as \emph{Audio-SME}. Vid-SME measures membership evidence through changes in model uncertainty under temporal reversal. To extend this principle to audio, we divide a recording $a$ into consecutive fixed-length chunks of duration $c$ and reverse their order, obtaining $a^{\mathrm{rev}}$. The samples within each chunk remain unchanged, preserving local acoustic content while disrupting the recording's global temporal structure.

We separately extract the model's audio-associated output distributions for $a$ and $a^{\mathrm{rev}}$, denoted by $p_t$ and $p_t^{\mathrm{rev}}$, respectively. Following the original Vid-SME implementation, we compute Sharma--Mittal entropy with fixed parameters $q=2$ and $r=1$, corresponding to second-order R\'enyi entropy. The position-wise entropy difference is
\begin{equation}
    \Delta_t(a)
    = H_2(p_t)-H_2(p_t^{\mathrm{rev}}).
\end{equation}
Audio-SME obtains an audio-level membership score by aggregating $\{\Delta_t(a)\}$ using the mean or the Min-K\% and Max-K\% aggregation functions provided by Vid-SME. We also evaluate the difference between sequence-level entropy values computed from the mean probability distribution across audio-associated positions. This adaptation retains Vid-SME's entropy-based scoring mechanism while replacing video-frame reversal with audio-chunk reversal.

\textbf{VideoMIA adaptation.}
We further adapt VideoMIA~\cite{videomia}, a response-only membership inference method that evaluates a model's ability to reason about temporal relationships. Given a recording $a$, we repeatedly construct temporal-ordering trials by sampling a sequence of consecutive context clips, followed by four non-overlapping candidate clips from later positions in the same recording. 

The candidate clips are randomly shuffled and assigned labels $\{\mathrm{A},\mathrm{B},\mathrm{C},\mathrm{D}\}$. For each trial $j$, we prompt the target model to perform two tasks: (i) \emph{next-segment identification}, which asks the model to identify the earliest candidate following the context; and (ii) \emph{temporal ranking}, which asks the model to arrange all four candidates in chronological order. Both tasks are evaluated against the original temporal positions of the sampled clips.

For identification, we assign a score
\begin{equation}
    u_j =
    \begin{cases}
        2^{-(d_j-1)}, & d_j\in\{1,2,3\},\\
        0, & \text{otherwise},
    \end{cases}
\end{equation}
where $d_j$ is the true chronological rank of the candidate selected by the model. For temporal ranking, we measure the proportion of correctly ordered candidate pairs:
\begin{equation}
    v_j =
    \frac{1}{\binom{4}{2}}
    \sum_{1\leq k<l\leq 4}
    \mathbb{I}\left[
        \hat{\pi}_j(k)<\hat{\pi}_j(l)
    \right],
\end{equation}
where $\hat{\pi}_j(k)$ denotes the predicted position of the candidate whose true chronological rank is $k$. Invalid or unparseable predictions receive a score of zero.
Finally, we aggregate the two task scores over $J$ independently sampled trials:
\begin{equation}
    S_{\mathrm{VideoMIA\text{-}Audio}}(a)
    =
    \frac{1}{2J}
    \sum_{j=1}^{J}(u_j+v_j).
\end{equation}
This response-only baseline assesses whether an audio LLM's temporal understanding provides evidence of recording-level training membership, without accessing token probabilities or training annotations.

\section{AudioTrace}
\label{sec:audiotrace}

\begin{figure*}[t]
    \centering
    \includegraphics[width=0.95\linewidth]{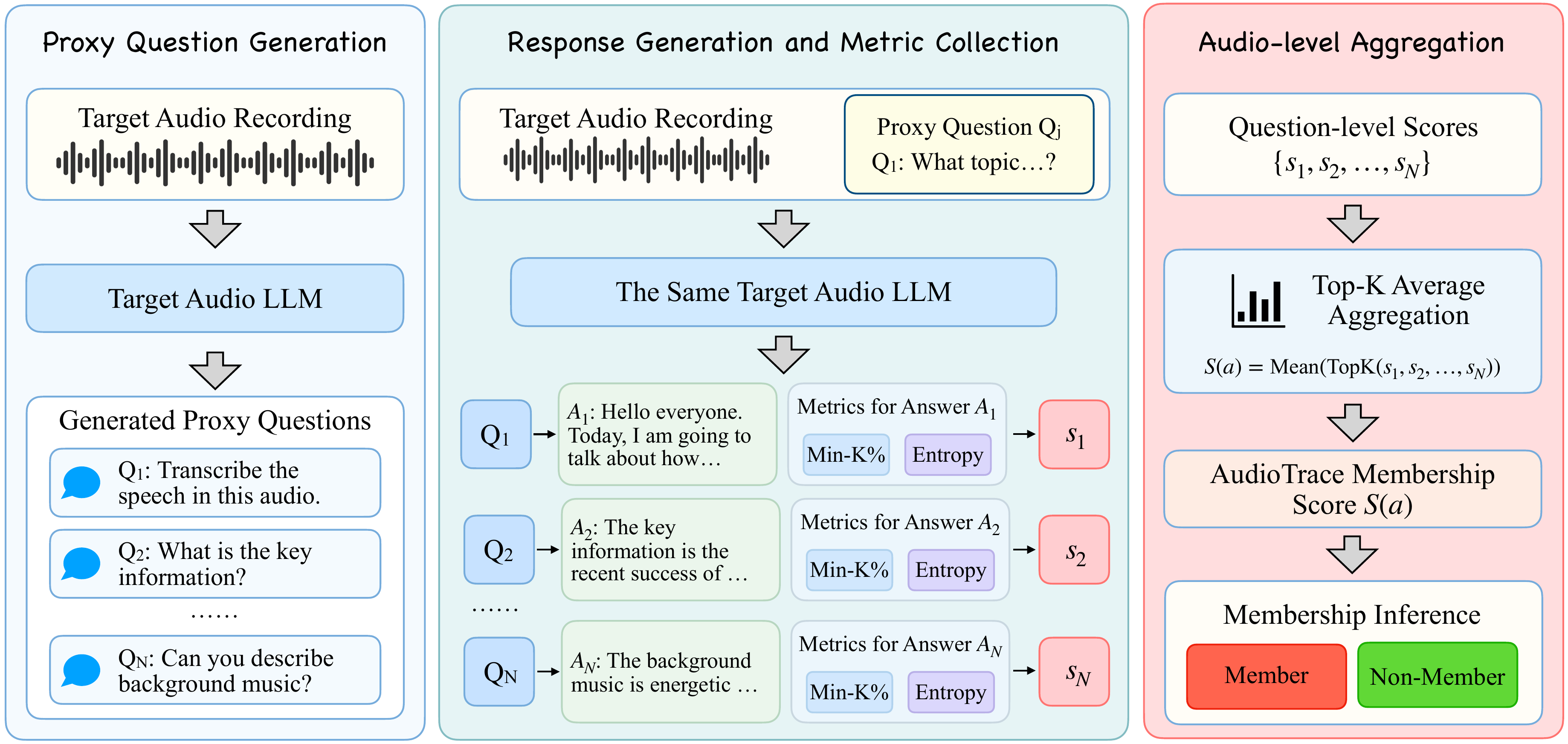}
    \caption{AudioTrace membership inference method. AudioTrace comprises of 3 stages from left to right. It first generates proxy questions from the target audio LLM. Then, for each proxy question, the target LLM generates an answer and the membership signals, such as Min-K\% and predictive entropy, are recorded. Finally, these signals are aggregated at audio-file-level using top-k average, producing one single membership score for each audio file.}
    \label{fig:audiotrace}
\end{figure*}

Direct audio-file-level MIAs attempt to detect membership signals from the output distributions of audio-associate positions or temporal dependencies. However, Audio LLMs are commonly fine-tuned to generate textual responses conditioned on both audio recordings and task-specific instructions. Consequently, the membership signal may be more apparent in the model's responses to relevant questions than in its direct audio representations. In practice, the original training instructions are generally unknown to the attacker, making conventional question-answer-based membership inference methods difficult to apply.

To address this challenge, we propose \textbf{AudioTrace}, a proxy-question-based approach to audio-file-level membership inference. As illustrated in Fig.~\ref{fig:audiotrace}, AudioTrace leverages the target audio LLM itself to construct questions about a candidate recording, obtains the target model's responses to those questions, and aggregates their membership scores into a single file-level prediction. The method consists of three stages: proxy-question generation, response generation and metric collection, and file-level aggregation. Importantly, no stage requires access to the original training annotations.

\subsection{Proxy-Question Generation}
\label{sec:proxy_question}

Membership signals in Audio LLMs may depend strongly on the probing instructions, but the original training questions are typically unavailable to the attacker. We therefore leverage the target model to generate proxy questions, hypothesising that its learned audio-instruction associations favour questions aligned with its training supervision, thereby eliciting stronger membership signals.
Specifically, given an audio recording $a$ and the target model $f_\theta$, AudioTrace first generates a set of $N$ proxy questions about the recording. We provide $a$ together with a fixed question-generation instruction $q_{\mathrm{gen}}$ to the target model:
\begin{equation}
    Q(a) = \{q_1,\ldots,q_N\}
    = \operatorname{Generate}_{f_\theta}(a,q_{\mathrm{gen}}).
\end{equation}
The instruction (Appendix \ref{sec:prompt}) encourages the model to formulate questions based on the audio content, potentially covering different aspects such as spoken content, speakers, and background sounds. We do not explicitly remove repeated or semantically similar questions, as their recurrence may reflect the model's preference for particular probing contexts. Such preferences may be informative for membership inference, though they themselves may not be a signal.

Unlike predefined transcription or captioning prompts, these proxy questions are conditioned on the individual recording and can probe multiple aspects of its content. Meanwhile, transcription and caption queries are also included in the set of proxy queries.

\subsection{Response Generation and Metric Collection}

We then query the \emph{same target model} with the original recording $a$ and each generated question $q_j$, obtaining a corresponding answer:
\begin{equation}
    \hat{y}_j =
    \operatorname{Generate}_{f_\theta}(a,q_j),
    \qquad j=1,\ldots,N.
\end{equation}
This produces a collection of audio-question-answer triplets
$\{(a,q_j,\hat{y}_j)\}_{j=1}^{N}$,
where both the questions and answers are generated by the target model, without requiring access to the training set annotations or reference responses.

For each generated answer $\hat{y}_j=(\hat{y}_{j,1},\ldots,\hat{y}_{j,T_j})$, we extract the target model's token-level output distributions conditioned on the audio recording and proxy question. Let
\begin{equation}
    p_{j,t}(v)
    =
    P_\theta\left(
        v \mid a,q_j,\hat{y}_{j,<t}
    \right)
\end{equation}
denote the probability of token $v$ at position $t$ of the generated answer. Unlike direct audio-position scoring, the generated response provides the actual tokens against which their probabilities can be evaluated.

We consider three families of membership statistics: Min-K\% probability, Max R\'enyi entropy and Shannon entropy. For Min-K\%, we select the fraction $\rho$ of answer tokens with the lowest log-probabilities. Let $\mathcal{T}_{j,\rho}$ denote the corresponding token positions. The question-level score is
\begin{equation}
    s_{j,\mathrm{MinK}}
    =
    - \frac{1}{|\mathcal{T}_{j,\rho}|}
    \sum_{t\in\mathcal{T}_{j,\rho}}
    \log p_{j,t}(\hat{y}_{j,t}).
\end{equation}
For entropy-based scoring, we compute the Shannon entropy and R\'enyi entropy with a given $\alpha$ of the predictive distribution $p_{j,t}$ at each answer position $t$, denoted as $H(p_{j,t})$.
The question-level entropy score is obtained by averaging over the generated answer as follows.
\begin{equation}
    s_{j,\mathrm{Ent}}
    =
    \frac{1}{T_j}
    \sum_{t=1}^{T_j}H(p_{j,t}).
\end{equation}
For R\'enyi entropy, we use the Max-K\% statistics by averaging across the fraction $\rho$ of the answer positions with the largest entropy values.
All scores are oriented such that larger values indicate higher uncertainty and, potentially, stronger non-member evidence of the given query-response pair, which aligns with the top-K aggregation next.

\begin{figure*}[t]
    \centering
    \includegraphics[width=0.8\linewidth]{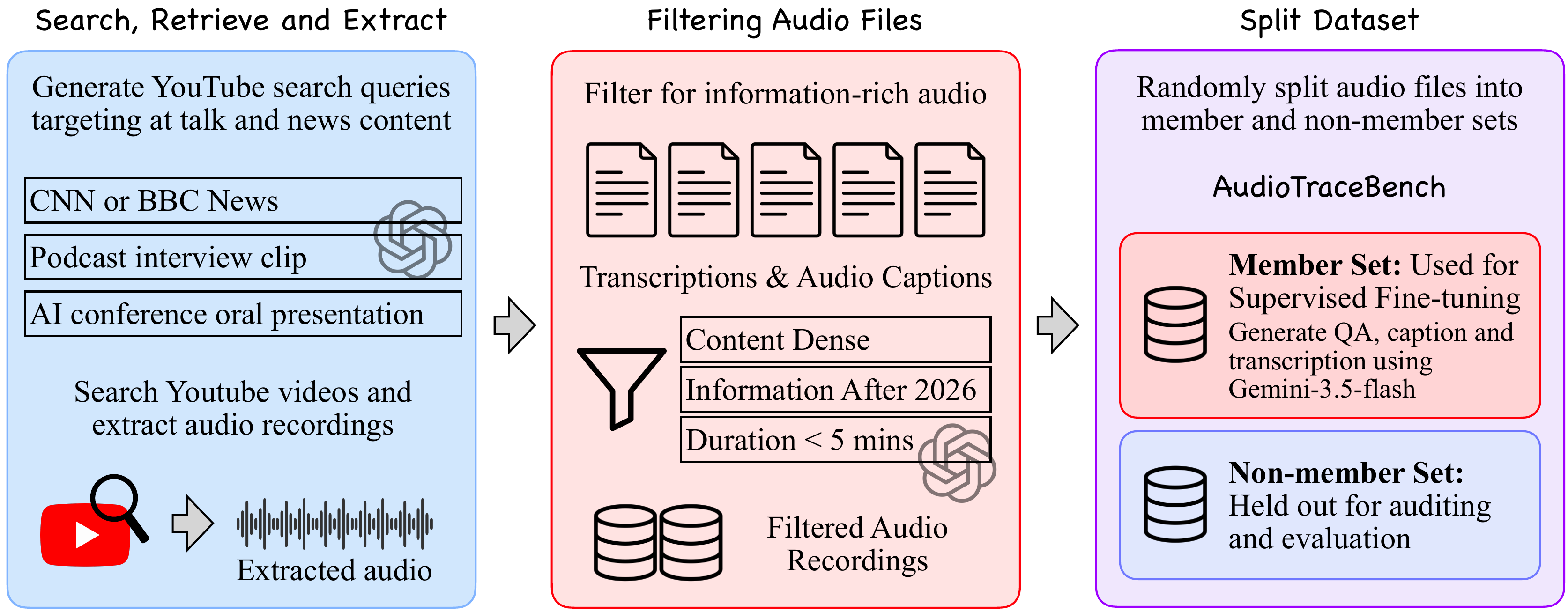}
    \caption{Data generation and curation pipeline for AudioTraceBench. We first collect YouTube videos from time-sensitive, information-rich categories and extract their audio tracks. We then filter the recordings based on content recency, information density, and duration. Finally, we divide the filtered recordings into a member set for supervised fine-tuning and a non-member set held out for membership auditing.}
    \label{fig:audiotracebench}
\end{figure*}

\subsection{File-level Aggregation}
\label{sec:file_aggregation}
Because different proxy questions may yield varying degrees of membership evidence, we aggregate their scores at the recording level rather than treating each question as an independent membership target. For a chosen metric $M\in\{\mathrm{MinK},\mathrm{Ent}\}$, let
$\{s_{j,M}\}_{j=1}^{N}$ denote its question-level scores. We adopt \emph{Top-$K$ average aggregation}, selecting the $k$ highest-scoring questions:
\begin{equation}
    S_M(a)
    =
    \frac{1}{k}
    \sum_{j\in\operatorname{TopK}_k
    (\{s_{i,M}\}_{i=1}^{K})}
    s_{j,M}.
\end{equation}
The resulting score $S_M(a)$ is used as the recording-level membership score $s_{\mathcal{A}}(a;f_\theta)$ defined in Sec.~\ref{sec:problem_formulation}. AudioTrace evaluates Min-K\% and entropy-based variants separately, allowing each metric to provide an independent audio-file-level membership estimate.
By aggregating evidence across multiple automatically generated questions, AudioTrace reduces its dependence on any single instruction and enables membership inference without knowing how the recording was annotated during fine-tuning.

\section{AudioTraceBench}
Reliable evaluation of audio-file-level MIA requires accurate membership labels and careful control of potential dataset biases. However, the complete training corpora of existing Audio LLMs are rarely disclosed, making it difficult to establish whether publicly available recordings were previously encountered by a model. To address this challenge, we construct \textbf{AudioTraceBench}, a controlled benchmark in which membership is explicitly determined through SFT. AudioTraceBench contains 1,285 unique recordings, randomly divided into 640 members and 645 non-members. Fig.~\ref{fig:audiotracebench} illustrates the construction pipeline.

\subsection{Data Creation Principle}

\begin{figure}[h]
    \centering
    \includegraphics[width=0.85\linewidth]{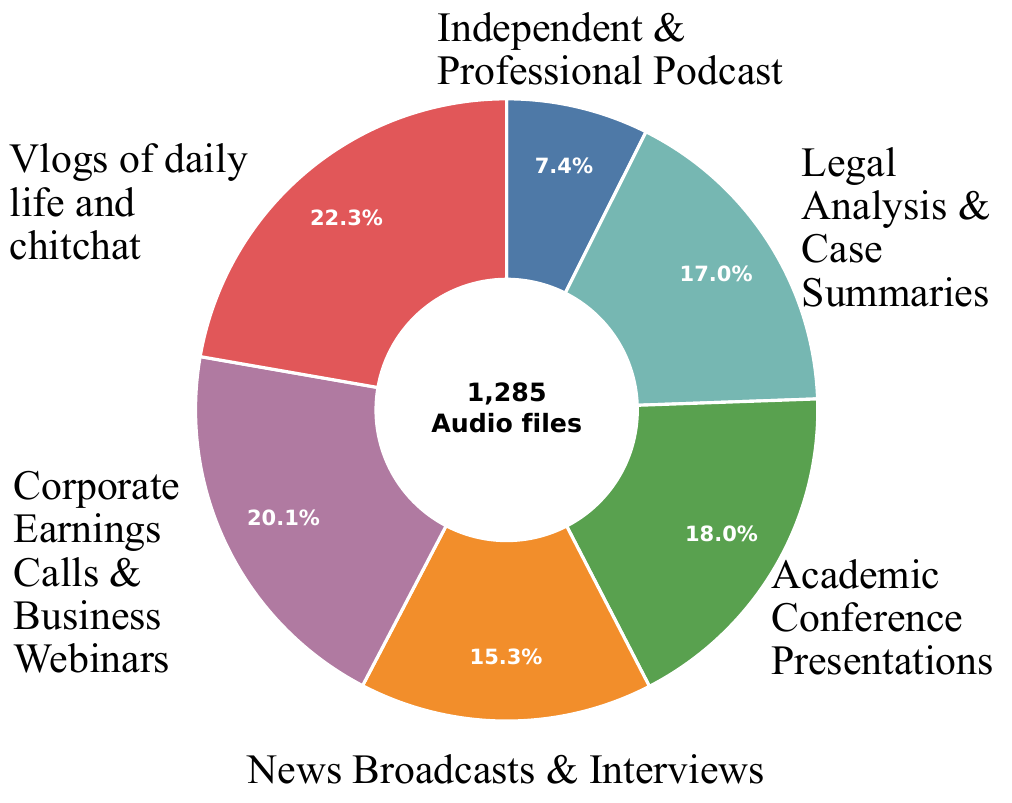}
    \caption{Pie chart showing the distribution of the number of audio files across six different audio categories in AudioTraceBench.}
    \label{fig:cat_dist}
\end{figure}


\noindent\textbf{Temporal separation.}
We collect recordings from YouTube videos published after January 1, 2026, and evaluate audio LLM checkpoints obtained before 2026. To further reduce potential overlap with the models' previous training data, we prioritize time-sensitive content, particularly news, interviews, and discussions referencing recent events. The categories of audio files and their distribution are provided in Fig. \ref{fig:cat_dist}. This temporal separation reduces the likelihood of prior exposure to the selected recordings, although publication dates alone cannot rule out reused content.

\noindent\textbf{Information-density filtering.}
After retrieving videos and extracting their audio, we apply captioning and transcription to assess the information content of each recording. Specifically, we decompose each generated caption into atomic facts and count the resulting factual statements. We retain recordings containing more than 20 atomic facts and more than 100 transcribed words, thereby favouring information-rich audio with substantial spoken content. We additionally restrict recording duration to at most five minutes to ensure compatibility with the evaluated Audio LLMs.

\noindent\textbf{Controlled membership assignment.}
After filtering, we randomly partition the recordings into disjoint member and non-member sets, containing 640 and 645 recordings, respectively. Only member recordings are used for SFT, while non-members remain held out for auditing. Since both sets are sampled from the same collection and filtering pipeline, this design reduces systematic differences in data provenance and content characteristics, while providing explicit membership ground truth.

\noindent\textbf{Multi-task annotation.}
For each member recording, we prompt Gemini 3.5 Flash to generate one transcription, one audio caption, and 5-10 question-answer (QA) pairs covering spoken content, speaker attributes, and background sounds, following practices adopted in many Audio LLMs \cite{salmonn,salmonn2,qwen2audio}. On average, each member recording is associated with 7.45 QA pairs. These annotations represent a typical SFT dataset for Audio LLMs and are set to be unavailable during membership inference.

\subsection{Dataset Statistics}

\noindent\textbf{Dataset composition.}
AudioTraceBench comprises 1,285 unique recordings, including 640 members (49.8\%) and 645 non-members (50.2\%). Fig.~\ref{fig:cat_dist} presents the distribution across six categories, which encompass diverse conversational settings, speaking styles, and acoustic environments.

\noindent\textbf{Duration and supervision statistics.}
Member recordings have a mean duration of $138.3 \pm 35.2$ seconds, compared with $136.7 \pm 35.0$ seconds for non-members, indicating closely matched duration distributions. 

\noindent\textbf{Blind membership evaluation.}
To assess whether membership labels can be inferred from pre-existing model knowledge rather than the controlled SFT intervention, we evaluate the original pretrained checkpoints before fine-tuning, i.e. $f_{\theta_0}$. Using available QA annotations, we compute perplexity, Min-30\% probability, and entropy-based membership scores. For Qwen2.5-Omni, the three methods achieve AUROCs of 0.4977, 0.4986, and 0.4977, respectively. The corresponding results for SALMONN-2 are 0.4930, 0.4975, and 0.4841. All results are approximately chance-level, indicating that these scoring methods do not readily distinguish the designated members from non-members before SFT.
\section{Experimental Setup}
\label{sec:exp_setup}

\subsection{Models and Fine-Tuning}
\label{sec:exp_models}

Our experiments cover three audio-capable LLMs, including Qwen2.5-Omni~\cite{qwen25omni}, SALMONN 2~\cite{salmonn2}, and Step-Audio-2-mini~\cite{stepaudio2}. All three models are at the 7B--8B parameter scale. For Qwen2.5-Omni and Step-Audio-2-mini, we use officially released checkpoints available before 2026 to establish temporal separation from AudioTraceBench. \xiao{For SALMONN 2, its fully public training data and pipeline allow us to verify that all backbone models were trained on pre-2026 data.} 
We initialise each model from its pretrained checkpoint and perform SFT using the member recordings in AudioTraceBench.

The training data for SFT contains audio captioning, speech transcription and QA, representing a typical fine-tuning setup. We adopt low-rank adaptation (LoRA)~\cite{hu2021lora} for all three models, with rank $r=128$ and scaling factor $\alpha=256$. LoRA is applied to linear layers in the self-attention block as well as the feed-forward networks in the LLM backbone, while keeping the audio encoder part frozen. We use a learning rate of $2\times10^{-5}$, an effective batch size of 8. Each model is fine-tuned for three epochs, unless specified in the results for ablation studies. Training of each model is performed on a single NVIDIA A100 GPU, with a three-epoch run taking approximately 20 hours.

\subsection{Inference Details and Baseline Setup}
\label{sec:exp_implementation}

For AudioTrace, we generate 30 proxy questions per recording using temperature sampling with $T=1$. Adding one transcription and one caption query, the total number of proxy questions is 32 for each audio file. Answers to the proxy questions are generated using greedy decoding. We evaluate top-$k$ aggregation with a default value of $k=2$ and use Entropy, Min-k\% and Max R\'enyi Entropy with token selection fractions of $\{30\%,60\%\}$. The corresponding configurations are specified in the results tables. For permutation-based methods such as Audio-SME and VideoMIA-Audio, we use an audio chunk size of 2 seconds. As a stronger baseline for membership inference, we design a \textit{fixed question bank} containing 30 questions that apply to all audio files (see Appendix for the questions). The same aggregation as AudioTrace across questions is applied to obtain a single membership signal.

Greedy decoding is always applied to answer both fixed questions and generated questions. All inference configurations are fixed across member and non-member recordings. The code for fine-tuning, inference and scoring is provided\footnote{\url{https://anonymous.4open.science/r/AudioMIA_Clean-DA3E/}}.

\subsection{Evaluation Protocol}
\label{sec:exp_eval}

Membership inference is performed at the audio-file level, using the same 640 member and 645 non-member recordings across all target models and attacks. Each recording contributes a single membership score, regardless of the number of associated SFT examples or proxy questions. For methods producing multiple scores per recording, we apply their respective aggregation strategies.
Unless explicitly designated as an analytical comparison, all attacks operate without knowledge of the original SFT questions or responses. We orient membership scores consistently such that higher values indicate stronger evidence of membership.

We report the area under the receiver operating characteristic curve (AUROC) and the true positive rate at a 5\% false positive rate (TPR@5\% FPR). AUROC evaluates overall discrimination between members and non-members, while TPR@5\% FPR measures detection performance under a low false-positive budget. Random guessing corresponds to an expected AUROC of 0.5 and TPR@5\% FPR of 0.05.
\section{Results}
\TakeawayBox{
Experiments are organised around four key questions to evaluate audio-file-level membership inference and investigate the factors influencing membership signals. (i) Can membership be inferred without training annotations? \S\ref{sec:mainresults-audiotracebench}-\S\ref{sec:blindbaseline} show that existing methods perform near chance, while AudioTrace achieves AUROCs of up to 0.60 across three Audio LLMs. Blind baselines further support the reliability of our benchmark. (ii) How do unknown supervision and audio content affect membership inference? \S\ref{sec:qa_perturbation} reveals a substantial performance gap between known and unknown training annotations, while perturbation experiments demonstrate the dependence of membership signals on acoustic content. (iii) Why do proxy questions matter?  \S\ref{sec:proxyq} shows that the generation and alignment of proxy questions with training supervision significantly influence membership detection. (iv) How do training configurations affect membership signals? \S\ref{sec:ablation-trainingconfiguration} examines the effects of training tasks, sample exposure, and adaptation capacity, showing stronger membership signals under increased exposure and adaptation capacity.
}

\subsection{Main Results on AudioTraceBench}
\label{sec:mainresults-audiotracebench}

\begin{table*}[t]
\centering
\caption{
Audio-file-level membership inference performance on
\textbf{AudioTraceBench}, where each model is fine-tuned on the member set of AudioTraceBench.
We report AUROC and TPR at 5\% FPR for all metrics. Audio-SME and VideoMIA-Audio are adaptations of existing video-level MIA methods to audio introduced in \S\ref{sec:direct_audio_mia}. 
}
\label{tab:main_results}

\setlength{\tabcolsep}{5pt}
\renewcommand{\arraystretch}{1.12}
\small
\resizebox{0.85\linewidth}{!}{%
\begin{tabular}{llcccccc}
\toprule
\multirow{2}{*}{\textbf{Attack Method}} &
\multirow{2}{*}{\textbf{Setting}} &
\multicolumn{2}{c}{\textbf{Qwen2.5-Omni}} &
\multicolumn{2}{c}{\textbf{SALMONN-2}} &
\multicolumn{2}{c}{\textbf{Step-Audio-2-mini}} \\
\cmidrule(lr){3-4}
\cmidrule(lr){5-6}
\cmidrule(lr){7-8}
& & AUROC & TPR@5\%FPR & AUROC & TPR@5\%FPR & AUROC & TPR@5\%FPR \\
\midrule
PPL &                    &0.4957 & 0.0419 & 0.4970 & 0.0403 & 0.5207 & 0.0636 \\
Entropy &        & 0.4953 & 0.0419 & 0.4958 & 0.0372 & 0.5200 & 0.0698 \\
\midrule
\multirow{3}{*}{\shortstack[l]{Min-K\%}}
& Min-10\%                & 0.4987 & 0.0279 & 0.4916 & 0.0388 & 0.5141 & 0.0589\\
& Min-30\%                & 0.4946 & 0.0481 & 0.4933 & 0.0310 & 0.5173 & 0.0667 \\
& Min-60\%                &  0.4959	& 0.0459 &	0.4959 &	0.0372 &	0.5204 &	0.0636 \\
\midrule
\multirow{6}{*}{\shortstack[l]{Max-K\% R\'enyi Entropy}}
& Max-10\% ($\alpha=0.5$) & 0.5000 & 0.0341 &	0.5001 &	0.0388 &	0.5119 &	0.0604 \\
& Max-30\% ($\alpha=0.5$) &0.4992	& 0.0512 &	0.4972 &	0.0341 &	0.5152 &	0.0589 \\
& Max-60\% ($\alpha=0.5$) & 0.4998	& 0.0465 &	0.4956 &	0.0326 &	0.5172 &	0.0651 \\
& Max-10\%  ($\alpha=2$) & 0.5013	& 0.0248 &	0.4944 &	0.0372 &	0.5137 &	0.0558 \\
& Max-30\% ($\alpha=2$) & 0.4952 & 0.0527 &	0.4925 &	0.0310 &	0.5168 &	0.0667 \\
& Max-60\% ($\alpha=2$) & 0.4965 & 0.0465 & 0.4947 &0.0372 & 0.5195	& 0.0636\\
\midrule
\multirow{3}{*}{\shortstack[l]{Audio-SME}}
& Audio-SME Max 10\%             & 0.5106 & 0.0264 &	0.5005 &	0.0419 &	0.5194 &	0.0465 \\
& Audio-SME Max 30\%             &0.5115 & 0.0512 & 0.5068 & 0.0434& 0.5214 & 0.0574 \\
& Audio-SME Max 60\%             & 0.5072 & 0.0372 & 0.5133 & 0.0481 & 0.5229 & 0.0620 \\
\midrule
\multirow{2}{*}{\shortstack[l]{VideoMIA-Audio}}
& Trial = 5       & 0.5027 & 0.0372& 0.5015 & 0.0403 & 0.5133 & 0.0562 \\
& Trial = 15       & 0.5049 & 0.0337 & 0.5184 & 0.0496 & 0.5179 & 0.0602 \\
\midrule
\multirow{3}{*}{\shortstack[l]{Fixed Question Bank}}
& Entropy & 0.5224 & 0.0295 & 0.5390 & 0.0620 & 0.5250 & 0.0576\\
& Min-60\% & 0.5199 & 0.0357 & 0.5440 & 0.0481 & 0.5213 & 0.0525 \\
& Max-60\% R\'enyi-2 & 0.5159 & 0.0341 & 0.5455 & 0.0574 & 0.5178 & 0.0469 \\
\midrule
\multirow{6}{*}{\textbf{AudioTrace} (ours)}
& Entropy     & 0.5505  & 0.0589 & 0.5809  & 0.0822 & 0.5897 & 0.0930 \\
& Min-30\%    & 0.5525  & \textbf{0.0589} & 0.5896  & 0.0961 & 0.5949  & \textbf{0.1054} \\
& Min-60\% & 0.5555 & 0.0512 &  0.5881 & 0.0977 & 0.5939  & 0.1008  \\
& Max-30\% R\'enyi-2     & 0.5537  & 0.0558 & \textbf{0.5917}  & 0.0946& \textbf{0.5958}  & 0.1039 \\
& Max-60\% R\'enyi-2 & \textbf{0.5563}  & 0.0558 & 0.5899 & \textbf{0.1023} & 0.5948  & 0.0977 \\
& Combined & 0.5515 & 0.0574 & 0.5902 & 0.0992 & 0.5952 & 0.1039 \\
\bottomrule
\end{tabular}}
\end{table*}

The main results on AudioTraceBench across three different models fine-tuned on the member set are shown in Table \ref{tab:main_results}. AUROC and TPR at 5\% FPR
are reported for direct audio-based baselines and AudioTrace.
We highlight the following observations from the experiments.

\textbf{Existing audio-level methods perform close to random guessing}. Across all three models, conventional probability-based methods, including PPL, entropy, Min-K\%, and Max-K\% R\'enyi entropy, achieve AUROCs close to 0.5. For example, the best AUROCs among these methods are 0.5013, 0.5001, and 0.5207 on Qwen2.5-Omni, SALMONN 2, and Step-Audio-2-mini, respectively. Compared to the conventional methods, the adaptation of Vid-SME and VideoMIA to Audio-SME and VideoMIA-Audio, respectively, achieved similarly limited improvements. The best Audio-SME setting achieved slightly better AUROC scores across the models. The best VideoMIA-Audio method merely improved upon random guessing and is worse than Audio-SME. The best-performing baseline is the fixed question bank, indicating the necessity of generation-based membership signal in audio-file-level auditing. However, all the baselines have TPRs at 5\% FPR generally below random.


\textbf{AudioTrace consistently improves audio-file-level membership inference}, achieving significantly higher performance than conventional methods and the adapted Audio-SME and VideoMIA-Audio. Specifically, AudioTrace achieves 4.4\%, 4.7\% and 7.1\% absolute AUROC improvements compared to the Fixed Question Bank baseline on Qwen2.5-Omni, SALMONN 2, and Step-Audio-2-mini, respectively.
At 5\% FPR, AudioTrace with combined metric achieves TPRs of 0.0574, 0.0992, and 0.1039, compared with 0.0295, 0.0620, and 0.0576 for the fixed question bank baseline. The improvement is particularly evident on SALMONN 2 and Step-Audio-2-mini, where AudioTrace approximately doubles the detection rate permitted by a 5\% false-positive budget relative to the Fixed Question Bank baseline. These findings demonstrate that probing a recording through self-generated proxy questions can reveal membership information that is difficult to recover from direct audio-based membership signals or a pre-defined question set. While the top 2 scores are used by default, AudioTrace is robust to the choice of top-K, as shown in the analysis in \S\ref{sec:ablation_topk} in the Appendix.

\textbf{AudioTrace performance is robust across different metrics.} As shown in Table \ref{tab:main_results}, while single best results were achieved by different metrics, their differences were generally small, and are all well above the baselines. To achieve a single robust metric for auditing, we combine Min-30\%, Min-60\%, Max 30\% R\'enyi-2 and Max 60\% R\'enyi-2 metrics to give a single combined score for membership inference. The combined score does not achieve the best performance, but achieves the best balance between AUROC and TPR@5\%FPR across the table. Although audio-file-level membership inference remains challenging, AudioTrace represents a meaningful step forward, demonstrating that membership signals can be recovered even without access to the original training annotations.

\begin{table}[h]
    \centering
    \caption{AudioTrace operating under black-box setting.}
    \begin{tabular}{lcc}
    \toprule
    Membership score     & AUROC & TPR@5\%FPR \\
    \midrule
    Combined & \textbf{0.5515} & 0.0574 \\
    ROUGE-L & 0.5253 & \textbf{0.0731} \\
    BERT score     &  0.5441 & 0.0592 \\
    \bottomrule
    \end{tabular}
    \label{tab:blackbox}
\end{table}

\textbf{Black-box membership inference with AudioTrace.} AudioTrace can operate in a black-box setting by drawing samples instead of directly measuring the token-level probabilities. Specifically, to validate the effectiveness under the black-box setting, eight samples were drawn per proxy question from the LLM, and the consistencies across the samples were measured pairwise using ROUGE-L \cite{rouge} and BERTScore \cite{bertscore}. The pairwise consistency scores are averaged across all responses and are used as the membership inference signal, where a higher score indicates membership since the response distribution tends to be more concentrated around the training set responses. The results are shown in Table \ref{tab:blackbox} for Qwen2.5-Omni, where similar performance is achieved without access to token probabilities.

\subsection{Blind Baseline}

\label{sec:blindbaseline}

To assess whether the random membership split introduces detectable differences between members and non-members, we evaluate three blind classifiers using features extracted independently of the target Audio LLMs. The first uses a two-dimensional feature vector consisting of the average pitch ($f_0$) and duration. The second uses 768-dimensional text representations obtained by mean-pooling BERT-base embeddings \cite{bert} of the transcriptions. The third uses 512-dimensional acoustic representations obtained by averaging the feature encoder outputs of wav2vec 2.0 \cite{wav2vec2}. For each feature type, we train a linear binary classifier of the form $\sigma(\mathbf{w}^{\top}\mathbf{x}+b)$, where $\sigma$ denotes the sigmoid function and $\mathbf{w}$ and $b$ are trainable parameters estimated by maximum likelihood. We use a random 50-50 train-test split of AudioTraceBench and report the test set performance.

\begin{table}[h]
    \centering
    \small
    \caption{Blind baselines based on audio and content features. Binary classifiers are trained on each feature.}
    
    \begin{tabular}{p{5cm}c}
    \toprule
    Method & Accuracy \\
    \midrule
    Pitch and Duration     & 51.25\% \\
    Text Embeddings     & 52.77\% \\
    Acoustic Embeddings & 53.32\% \\
    \bottomrule
    \end{tabular}

    \label{tab:blindbaseline}
\end{table}

As shown in Table~\ref{tab:blindbaseline}, all three features achieve accuracies close to the 50\% chance level, suggesting that the tested audio and content features provide limited evidence for distinguishing members from non-members. This supports the effectiveness of our randomised membership assignment in reducing readily detectable dataset-level confounding.

\subsection{Challenge of Unknown Supervision Data}
\label{sec:qa_perturbation}

\begin{table*}[t]
    \centering
    \caption{Comparison of known training set QA annotations against unknown annotation condition for Qwen2.5-Omni and SALMONN-2 models under corruptive noise and audio tokeniser reconstruction perturbations. AudioTrace is used across the table with the standard setting as before.} 
    
    \begin{tabular}{p{7cm}ccccc}
\toprule
\multirow{2}{*}{\textbf{Setting}} &
\multirow{2}{*}{\textbf{Metric}} &
\multicolumn{2}{c}{\textbf{Qwen2.5-Omni}} &
\multicolumn{2}{c}{\textbf{SALMONN-2}} \\
\cmidrule(lr){3-4}
\cmidrule(lr){5-6}
& & AUROC & TPR@5\%FPR & AUROC & TPR@5\%FPR \\
\midrule
\rowcolor{SectionGray} AudioTrace - Known Training Set QA for Member Audio & & & & & \\
\multirow{3}{*}{Original Audio} & Min-60\% & 0.8104  & 0.4106 & 0.8216 & 0.4044 \\
& Max-60\% R\'enyi-2 & 0.8092  & 0.4121& 0.8206  & 0.3810\\
& Combined & 0.8152 & 0.4401 & 0.8234 & 0.4059 \\
\midrule
\multirow{3}{*}{Original Audio + 0dB SNR Noise} & Min-60\% & 0.5788 & 0.0529 & 0.5850  & 0.0871\\
& Max-60\% R\'enyi-2 & 0.5786  & 0.0529 & 0.5853  & 0.0731\\
& Combined & 0.5818 & 0.0451 & 0.5851 & 0.0855 \\
\midrule
\multirow{3}{*}{DAC-Reconstructed Audio} & Min-60\% & 0.8096   & 0.4077 & 0.8233 & 0.4122 \\
& Max-60\% R\'enyi-2 &  0.8085 & 0.4118 & 0.8179 & 0.3834 \\
& Combined & 0.8133 & 0.4379 & 0.8215 & 0.4073 \\
\midrule
\rowcolor{SectionGray} AudioTrace - Generated Proxy Questions & & & & & \\
\multirow{3}{*}{Original Audio} & Min-60\% & 0.5555 & 0.0512 &  0.5881 & 0.0977 \\
& Max-60\% R\'enyi-2 &  0.5563 & 0.0558 & 0.5899 & 0.1023  \\
& Combined & 0.5515 & 0.0574 & 0.5902 & 0.0992  \\
\midrule
\multirow{3}{*}{Original Audio + 0dB SNR Noise} &Min-60\% & 0.5037 & 0.0574 & 0.5221  & 0.0543\\
& Max-60\% R\'enyi-2 &  0.5113  & 0.0574 & 0.5228 & 0.0667\\
& Combined & 0.5061 & 0.0543 & 0.5216 & 0.0667 \\
\midrule
\multirow{3}{*}{DAC-Reconstructed Audio} & Min-60\% & 0.5545 & 0.0581 & 0.5881  & 0.0977\\
& Max-60\% R\'enyi-2 & 0.5553 & 0.0581 & 0.5898  & 0.1023\\
& Combined & 0.5532 & 0.0596 & 0.5903 & 0.0992 \\
\bottomrule
    \end{tabular}
    \label{tab:trainset}
\end{table*}

To investigate the source of membership signals, AudioTrace was compared against a privileged setting in which the original training questions and answers are known and are used the same way as how AudioTrace uses generated proxy questions. To investigate the association with the specific audio file, we further apply two audio perturbations, including highly destructive additive noise at 0\,dB SNR and reconstruction using a discrete audio codec (DAC). The former substantially corrupts the acoustic signal to check if the membership signal is tied to specific text query-response pairs, while the latter modifies the waveform but retains the same auditory content when listening to it. Table~\ref{tab:trainset} presents the results on Qwen2.5-Omni and SALMONN-2.

\textbf{Knowledge of training QA substantially strengthens membership inference.} On the original recordings, the privileged setting achieves AUROCs of up to around 0.82 for both Qwen2.5-Omni and SALMONN-2, respectively, with TPR@5\% FPR reaching over 0.40 for both models. In contrast, without accessing the original training annotations, AudioTrace only achieves AUROCs of at most 0.56 and 0.59, with corresponding maximum TPRs of 0.06 and 0.10. This substantial gap indicates that the original question-answer pairs provide strong membership evidence that is difficult to recover when only the audio recording is available. Consequently, performance under known training QA supervision should not be interpreted as representative of practical audio-file-level membership inference.

\textbf{Strong acoustic corruption largely eliminates AudioTrace's membership signal.} Adding noise at 0\,dB SNR markedly reduces attack performance in both settings. For the privileged setting, the best AUROC decreases drastically, from 0.82 to 0.58 on Qwen2.5-Omni and from 0.82 to 0.59 on SALMONN-2. Nevertheless, a residual membership signal remains, with AUROCs consistently above 0.57 across the evaluated metrics. This residual signal is consistent with the model retaining some level of information about the specific textual question-answer pairs encountered during training, even when the audio is heavily corrupted.

The effect is more evident for AudioTrace. Its best AUROC decreases from 0.56 to 0.51 on Qwen2.5-Omni and from 0.59 to 0.52 on SALMONN-2, approaching random guessing performance. This finding suggests that AudioTrace's membership evidence depends substantially on the acoustic information in the candidate recording. Unlike the privileged setting, AudioTrace cannot directly exploit knowledge of the original verbatim training questions and answers. The disappearance of its signal under severe audio corruption therefore provides evidence that AudioTrace probes recording-level membership, rather than merely detecting memorised textual question-answer pairs.

\textbf{Audio reconstruction preserves membership evidence despite waveform changes.} In contrast to additive noise, DAC reconstruction has little effect on both settings. For AudioTrace, the best AUROCs on reconstructed audio are 0.56 and 0.59 on Qwen2.5-Omni and SALMONN 2, compared with 0.57 and 0.59 on the original recordings. Similarly, the privileged setting maintains comparable AUROCs.

These results indicate that AudioTrace does not rely solely on an exact waveform match. Its membership signal survives DAC reconstruction, suggesting that relevant audio information is preserved. The perturbation experiments suggest that AudioTrace captures audio-dependent membership evidence that is robust to moderate reconstruction changes but vulnerable to severe acoustic corruption.

\subsection{Ablation Study on Proxy Questions}
\label{sec:proxyq}

Proxy questions can be generated either by the target model itself or by an external model. AudioTrace adopts the former approach, motivated by the hypothesis that a target model's generated questions better reflect the knowledge acquired during training. To examine this design choice, we use Qwen3-Omni-30B-A3B-Instruct \cite{qwen3omni} as an external question generator, producing 30 proxy questions per recording, and apply the same AudioTrace scoring and aggregation procedure. We further quantify the affinity between proxy questions and the original training questions using rescaled BERTScore \cite{bertscore}. Specifically, for each proxy question, we compute its BERTScore against every training question associated with the same recording and take the maximum similarity as its affinity score. We then average these scores across proxy questions and report the resulting affinity alongside membership inference performance, measured using the Combined score, in Table~\ref{tab:sourceofqa}.

\begin{table}[h]
    \centering
    \caption{Similarity to the training set questions and influence of the source of queries on auditing performance. The combined score is used to measure AUROC. We adopt the Qwen3-Omni-30BA3B-Instruct model to generate proxy questions as a comparison reference.}
    \begin{tabular}{lccc}
    \toprule
    Model & Source of Query     &  Similarity & AUROC \\
    \midrule
    Qwen2.5-Omni & self-gen. & 0.4109 & 0.5515 \\
    Qwen2.5-Omni & Qwen3-Omni gen. & 0.3870 & 0.5383 \\
    SALMONN-2 & self-gen.    & 0.4166 & 0.5902 \\
    SALMONN-2 &  Qwen3-Omni gen.   & 0.3870 & 0.5562  \\
    \bottomrule
    \end{tabular}
    \label{tab:sourceofqa}
\end{table}

As shown in Table~\ref{tab:sourceofqa}, self-generated proxy questions consistently achieve higher similarity to the original training questions and better membership inference performance than questions generated by an external model. For Qwen2.5-Omni, self-generated questions achieve a similarity score of 0.41 and an AUROC of 0.55, compared with 0.39 and 0.54 using Qwen3-Omni-generated questions. A similar trend is observed for SALMONN-2, where self-generated questions improve similarity from 0.39 to 0.42 and AUROC from 0.57 to 0.59. These results support our hypothesis that questions generated by the target model are more closely aligned with its training supervision, making them more effective for eliciting membership signals. 

\textbf{Influence of proximity between training questions and proxy questions.} The consistent association between higher question similarity and improved AUROC in Table \ref{tab:sourceofqa} suggests that proximity to the training-question distribution is an important factor. This finding further motivates the design choice of using the target model itself for proxy-question generation in AudioTrace. To further establish this correlation, we apply AudioTrace only to the $M$ most similar proxy queries and measure the performance variation against different $M$, with results shown in Table \ref{tab:similarity_topk}.

\begin{table}[t]
    \centering
    \caption{AUROC of different models using the $M$ most similar proxy queries to the training set questions.}
    \begin{tabular}{lccc}
    \toprule
    Model     & $M$ & Max-60\% R\'enyi-2 & Combined \\
    \midrule
    \multirow{5}{*}{Qwen2.5-Omni} & 3 & 0.5995 & 0.6209 \\
     & 5 & 0.5791 & 0.5973 \\
    & 10 & 0.5527 & 0.5669 \\
    & 20 & 0.5568 & 0.5527 \\
    & 30 & 0.5563 & 0.5515 \\
    \midrule
    \multirow{5}{*}{SALMONN-2} & 3 & 0.6778 & 0.6822 \\
    & 5 & 0.6253 & 0.6285 \\
    & 10 & 0.6179 & 0.6212 \\
    & 20 & 0.5959 & 0.5983 \\
    & 30 & 0.5899 & 0.5902 \\
    \bottomrule
    \end{tabular}
    \vspace{-0.3cm}
    \label{tab:similarity_topk}
\end{table}

The results in Table \ref{tab:similarity_topk} reveal that restricting AudioTrace to proxy questions more similiar to 
the original training questions substantially improves membership inference. With only the three most similar questions ($M=3$), AudioTrace achieves its highest AUROCs of 0.62 and 0.68 using the Combined score on Qwen2.5-Omni and SALMONN-2, respectively. As $M$ increases to 30, the AUROCs decrease to 0.55 and 0.59, matching the results in the main table. A similar trend is observed with Max-60\% R'enyi-2 scoring, particularly on SALMONN-2. These findings further support the relationship between proxy-question affinity and membership inference effectiveness, where questions semantically closer to the original training supervision tend to reveal stronger membership signals, while incorporating less similar questions dilutes the evidence. 

Results in Table \ref{tab:sourceofqa} and \ref{tab:similarity_topk} highlight the importance of instruction alignment in audio-file-level membership inference. Questions more closely aligned with the original training supervision tend to elicit stronger membership signals, while target-model-generated questions exhibit higher training-question similarity than externally generated ones. This supports our motivation for leveraging the target model itself to approximate informative probing instructions when the original training annotations are unavailable.

\begin{figure}[t]
    \centering
    \hspace{-0.5cm}
    \includegraphics[width=0.85\linewidth]{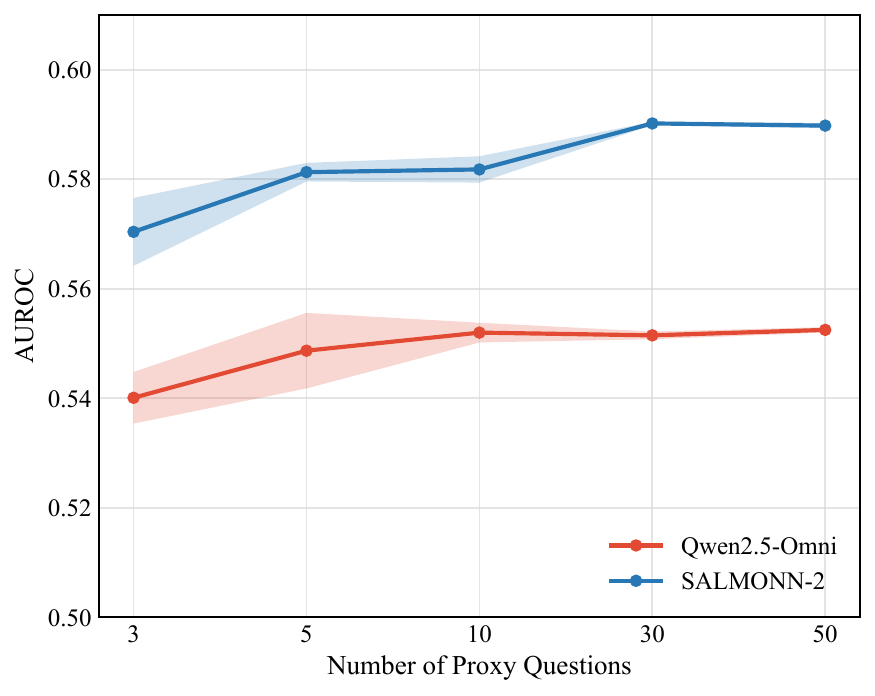}
    \caption{Variation of AUROC against the number of proxy questions generated for Qwen2.5-Omni and SALMONN-2 using AudioTrace Combined score. The shaded area represents the single standard deviation range.}
    \label{fig:questioncount}
\end{figure}

\begin{figure}[t]
    \centering
    \includegraphics[width=0.9\linewidth]{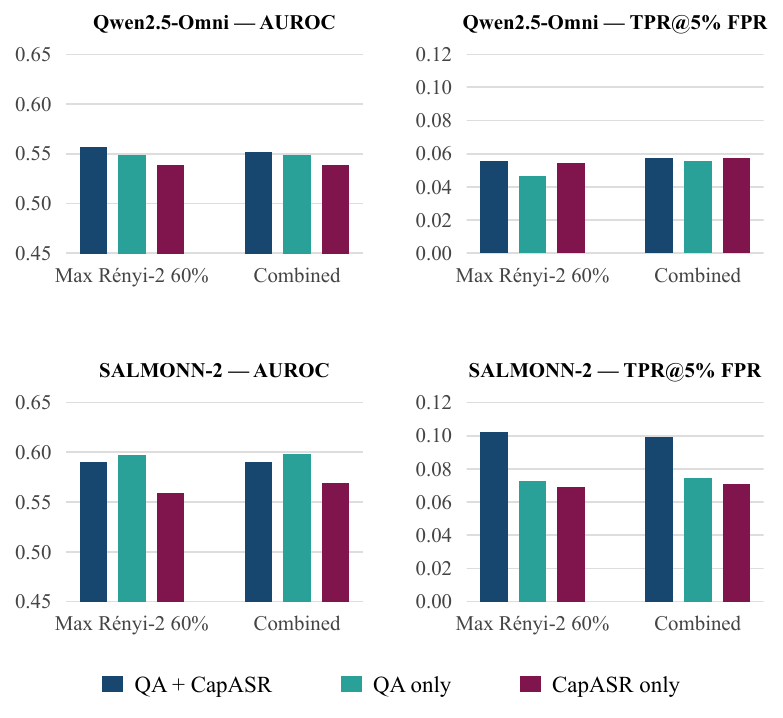}
    \caption{Variation of AUROC and TPR@5\%FPR with different training data composition, including QA, CapASR (caption and transcription) or the combination of the two, which is the default setting.}
    \label{fig:task_ablation}
\end{figure}

\begin{figure*}[t]
    \centering
    \includegraphics[width=0.9\linewidth]{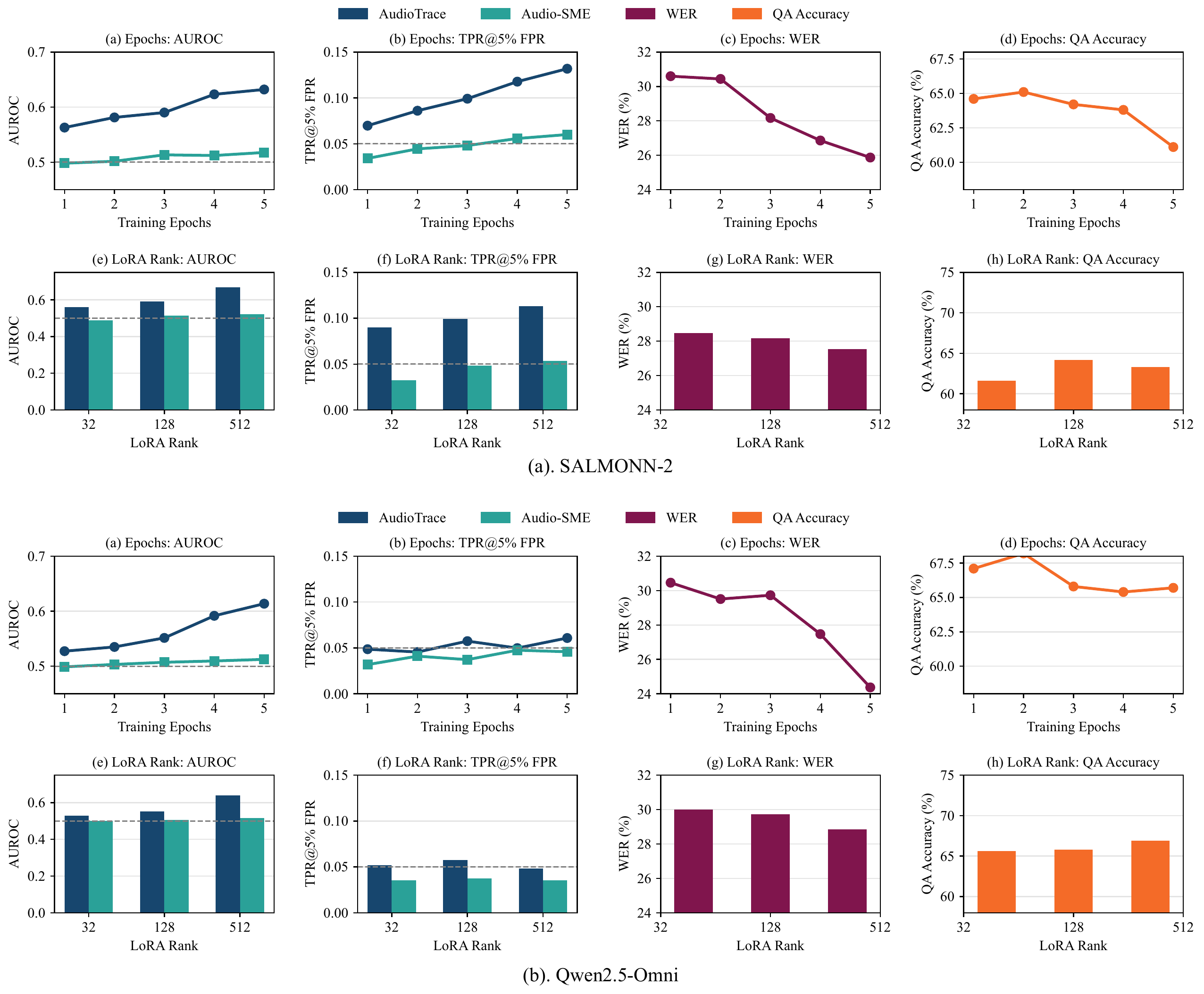}
    \vspace{-0.2cm}
    \caption{Ablation studies on training configurations, including the number of repetitions of the samples (represented by the number of epochs), and the number of trainable parameters controlled by the LoRA rank, using (a) SALMONN-2 and (b) Qwen2.5-Omni.}
    \label{fig:train_ablation}
\end{figure*}

\begin{table}[h]
    \centering
    \scriptsize
        \caption{Impact of composition of the proxy questions in AudioTrace measured by AUROC across all three models. The combined score is used as the membership score. For caption + ASR, the average of both tasks is used.}
    \begin{tabular}{lccc}
    \toprule
    Composition     & Qwen2.5-Omni & SALMONN-2 & Step-Audio-2-mini \\
    \midrule
    Caption + ASR     & 0.5276 & \textbf{0.6037} & 0.5549 \\
    QA & 0.5450 & 0.5723 & 0.5845 \\
    All & \textbf{0.5515} & 0.5902 & \textbf{0.5952} \\
    \bottomrule
    \end{tabular}
    \label{tab:proxy_composition}
\end{table}

\textbf{Impact of the number of proxy questions}. We investigate the effect of the number of generated proxy questions on AudioTrace's membership inference performance, as shown in Fig.~\ref{fig:questioncount}. For both Qwen2.5-Omni and SALMONN 2, AUROC generally improves as the number of proxy questions increases, but the gains gradually diminish when more samples are used. Specifically, increasing the question count from 3 to 50 improves AUROC from 0.54 to 0.55 on Qwen2.5-Omni and from 0.57 to 0.59 on SALMONN-2. Performance largely stabilises beyond 10 questions for Qwen2.5-Omni and 30 questions for SALMONN-2, suggesting diminishing returns from generating additional questions. Moreover, the standard deviation generally decreases as more questions are generated, indicating greater stability as the number of sampled questions increases because more samples cover the possible question space more thoroughly.

\textbf{Impact of proxy question composition.} Table~\ref{tab:proxy_composition} evaluates the contribution of different proxy-question types. Combining QA, captioning, and ASR achieves the highest AUROC on Qwen2.5-Omni and Step-Audio-2-mini, while remaining competitive on SALMONN 2 where captioning + ASR alone performs the best. The results suggest that different question types expose complementary membership evidence. By incorporating all three tasks, AudioTrace avoids relying on a single type of audio understanding and achieves consistently competitive performance without model-specific question selection, supporting our multi-task proxy-question generation design.

\subsection{Ablation Study on Training Configurations}

\label{sec:ablation-trainingconfiguration}

\textbf{Influence of training data composition.} To investigate how different composition in training data influences the performance of membership inference with AudioTrace method, Qwen2.5-Omni and SALMONN-2 were trained again just with audio captioning and speech recognition tasks, or just with QA. Other training configurations remained the same, and both models were trained for 3 epochs regardless of the training data composition. The performance of AudioTrace on those models are shown in Fig. \ref{fig:task_ablation}.

As shown in the figure, AudioTrace achieves higher membership inference performance when QA supervision is included alongside captioning and ASR during fine-tuning. When restricted to captioning and ASR, AudioTrace still outperforms random guessing and direct audio-based baselines, but with reduced effectiveness. One possible explanation is that membership evidence becomes concentrated in captioning- and ASR-related queries, while less informative QA queries dilute the aggregated signal. 
In this setting, focusing on captioning and ASR queries may yield stronger membership evidence. However, the exact composition of the training data is generally unavailable to the auditor, making such task-specific selection impractical. This further motivates AudioTrace's audio-file-level threat model, where membership must be inferred without knowledge of the training queries or supervision types.

We then investigate the influence of training exposure and adaptation capacity on audio-file-level membership inference, as shown in Fig.~\ref{fig:train_ablation}. We vary the number of training epochs, reflecting repeated exposure to member recordings, and the LoRA rank, which controls the number of trainable parameters.
To assess downstream utility, word error rate (WER) is measured on all 645 non-member recordings. Audio QA accuracy is evaluated on 2,000 question-answer pairs generated from these recordings by Gemini 3.5 Flash. Manual verification of 100 sampled questions yields 96\% annotation accuracy, and Qwen3.5-9B is used as an LLM judge to assess answer correctness.

\textbf{Impact of sample repetitions in training.} The AUROC of AudioTrace increases with increasing training epochs on both models, whereas Audio-SME remains close to random guessing. On SALMONN-2, the AUROC of AudioTrace increases from 0.56 to 0.63 over 5 epochs. Meanwhile, WER improves steadily, while QA accuracy declines with additional epochs. Similar improvements were found on Qwen2.5-Omni. These findings suggest that repeated exposure to member recordings strengthens the membership signal accessible through AudioTrace, even when downstream task performance exhibits different trends across tasks.

\textbf{Impact of LoRA rank.} Increasing the LoRA rank also strengthens the membership signal captured by AudioTrace. On SALMONN-2, AUROC increases from 0.56 to 0.67 as the rank increases from 32 to 512, while Audio-SME still remains substantially closer to random guessing. Similar improvements are observed on Qwen2.5-Omni. Notably, both WER and QA accuracy improve with increasing LoRA rank, indicating that stronger membership leakage can accompany improved downstream utility rather than necessarily resulting from degraded generalization.

\section{Discussion}

In this section, the broader implications of the findings are discussed for audio-file-level membership inference and the limitations of the proposed AudioTrace approach.

\textbf{Membership inference beyond fixed scoring metrics.} Existing membership inference research has primarily investigated how to extract membership evidence through prediction-based statistics~\cite{carlini2022membership,shi2024detecting,dong2026membership}. Our findings highlight another important dimension: how the model is queried. The substantial gap between known- and unknown-supervision settings, alongside the relationship between proxy-question similarity and attack performance, suggests that membership evidence is sensitive to probing context. This observation connects to broader work on data-use auditing under limited knowledge of model training~\cite{huang2024general}. It also complements document-level membership inference research~\cite{meeus2024did}, highlighting the importance of aligning membership auditing with the unit of interest to data owners. Our results motivate further research on informative query selection when training-time details are unknown.

\textbf{From membership inference to reliable auditing.} Prior security research has emphasized that successful membership inference does not necessarily provide reliable evidence of training data use. Carlini et al.~\cite{carlini2022membership} highlight the importance of evaluating attacks at low false-positive rates, while recent work examines the reliability and coverage of membership inference for privacy auditing~\cite{wang2025membershiptools}. Distributional confounding is another challenge, particularly for audio data~\cite{maini2024llm,dong2026membership}. Our controlled benchmark and blind baselines aim to mitigate these factors. Nevertheless, AudioTrace's modest AUROC and low-FPR detection performance indicate that its scores should be interpreted as statistical evidence rather than definitive proof of recording membership. Developing more reliable and well-calibrated membership evidence is therefore essential for practical privacy and copyright auditing.

\textbf{Limitations and Future Work.} Our study has several limitations. 
First, AudioTrace introduces a trade-off between inference efficiency and detection performance. Generating and evaluating multiple proxy questions requires approximately five times more inference computation than direct audio-level methods, but achieves higher AUROC. This additional cost may limit scalability in large-scale auditing scenarios. Second, our evaluation focuses on controlled supervised fine-tuning, while real-world training may involve larger, more diverse datasets, different supervision strategies, and varying sample exposure. Evaluating AudioTrace in such settings remains important. Finally, identifying effective proxy questions without knowledge of training details remains challenging. Future work could explore adaptive query selection to improve detection, particularly at low false-positive rates.
\section{Conclusion}

We investigate audio-file-level membership inference against Audio LLMs without knowledge of the original training queries and annotations. We introduce AudioTraceBench, a controlled benchmark revealing the limitations of existing MIA methods, and propose AudioTrace to generate audio-conditioned proxy questions and aggregate response-based membership signals at the file level. AudioTrace consistently outperforms adapted baselines across three Audio LLMs. Further analyses reveal how unknown training annotations, query selection, and fine-tuning configurations influence membership leakage. These findings lay the foundation for audio-file-level auditing and motivate more reliable membership inference methods.

\bibliographystyle{IEEEtran}
\bibliography{references}

\section*{Ethics Considerations}

Our study aims to support privacy and copyright auditing of Audio LLMs through membership inference. AudioTraceBench is constructed exclusively from publicly accessible YouTube videos. We screened the collected recordings and excluded those containing sensitive personal information to minimize potential privacy risks. To further mitigate privacy and copyright concerns, we will release only the original YouTube video links. Access to the benchmark through Hugging Face will be request-based.

We also acknowledge that AudioTrace could potentially be misused to infer the membership of recordings without their owners' authorization. However, our experiments are conducted using locally deployed models under controlled fine-tuning conditions, without targeting third-party production services. Furthermore, the attack requires access to token-level output probabilities and currently offers limited reliability for individual-recording decisions. We believe the benefits of understanding and mitigating training-data privacy risks justify studying this attack under these controlled conditions. Our study does not identify a specific vulnerability in a deployed service, and therefore no vendor-specific vulnerability disclosure has been undertaken.

\section{Appendix}
\subsection{Ablation on Aggregation Top-K}

\label{sec:ablation_topk}
\begin{figure}[h]
    \centering
    \includegraphics[width=0.99\linewidth]{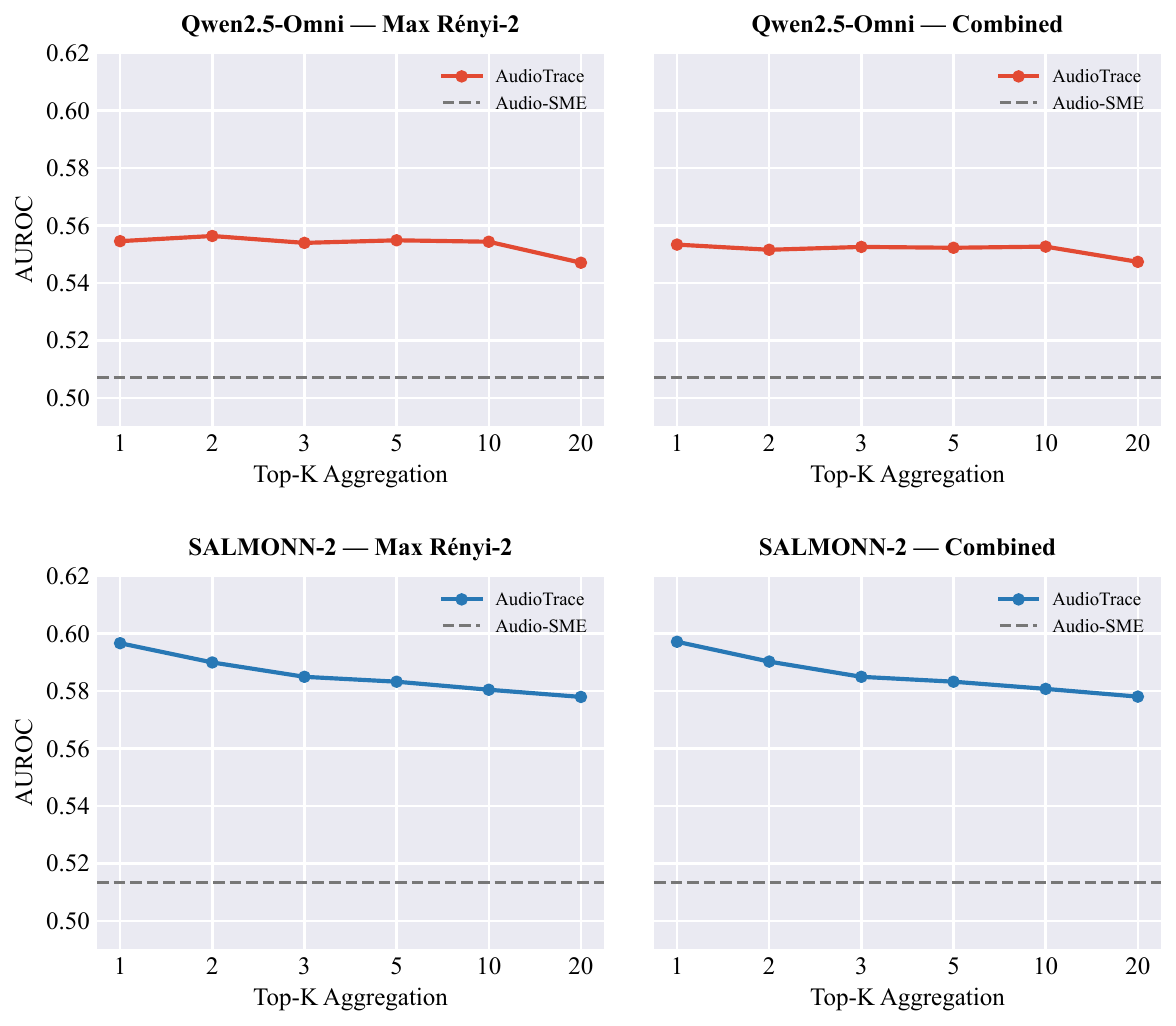}
    \vspace{-0.3cm}
    \caption{Variations of AUROC against different top-$K$ aggregations on AudioTraceBench on Qwen2.5-Omni and SALMONN-2 using Max-60\% R\'enyi-2 Entropy and the combined score.}
    \label{fig:topk}
\end{figure}

We investigate the effect of the top-$K$ aggregation parameter in Fig.~\ref{fig:topk}. Across both Qwen2.5-Omni and SALMONN 2, AudioTrace consistently outperforms the Audio-SME baseline over all evaluated values of $K$, indicating limited sensitivity to the aggregation parameter. Nevertheless, performance generally declines as more proxy questions are included, with absolute AUROC decreasing by approximately 1\% and 2\% on Qwen2.5-Omni and SALMONN 2, respectively. 

One plausible explanation is that some proxy questions may be less aligned with the training question distribution and therefore yield weaker membership signals. Including less informative questions in the aggregation may dilute the stronger evidence provided by the most revealing questions, suggesting that aggregating a small number of high-scoring questions is a preferable approach. %

\subsection{Fixed Question Bank}

\begin{lstlisting}[
    language=Python,
    basicstyle=\ttfamily\small,
    breaklines=true,
    frame=single,
    numbers=none,
    backgroundcolor=\color{gray!5}
]
"What is the main topic discussed in the audio?",
"What is the main purpose of the recording?",
"What is the most important information conveyed in the audio?",
"What are the key points discussed in the recording?",
"What is the main conclusion or takeaway?",
"What specific issue or problem is being discussed?",
"What event or situation is described in the audio?",
"What are the main facts presented in the recording?",
"What specific details are provided about the main topic?",
"What examples are mentioned to illustrate the discussion?",
"Who are the people mentioned in the audio?",
"What organizations or companies are mentioned?",
"What locations or places are mentioned?",
"What dates or time references are mentioned?",
"What numbers, quantities, or statistics are mentioned?",
"What happened at the beginning of the recording?",
"What events or developments are described later?",
"What outcome or result is discussed?",
"What changes or developments are mentioned?",
"What future plans or predictions are discussed?",
"What is the main argument or claim made by the speaker?",
"What reasons are given to support the main argument?",
"What evidence is provided to support the claims?",
"What concerns or challenges are raised?",
"What solutions or recommendations are proposed?",
"What opinions or viewpoints are expressed?",
"Are there any disagreements or contrasting perspectives?",
"What questions are asked during the recording?",
"What information does the speaker emphasize?",
"What uncertainties or limitations are mentioned?"
\end{lstlisting}

\subsection{Question-Generation Prompt}
\label{sec:prompt}
The following prompt is used for the target model to generate proxy questions in AudioTrace.

\begin{promptbox}[label={real_scripts}]{Question-Generation Prompt}
Generate a question about the audio. The question can be about the content, the speaker or background audio information. Output the question directly.\\
Question:
\end{promptbox}

\end{document}